\documentstyle[preprint,eqsecnum,aps]{revtex}

\def\ovl{\overline}
\def\bra{\langle}
\def\ket{\rangle}
\def\eqdef{\stackrel{\rm def}{\equiv}}
\def\dspl{\displaystyle}

\begin{document}
\preprint{HUTP-97/A021}
\title{Inclusive Decay Distributions of Coherent Two-body States}
\author{Hitoshi Yamamoto}
\address{Dept. of Physics, Harvard University, 
42 Oxford St., Cambridge, MA 02138, U.S.A.}
\date{\today}
\maketitle

\draft
\begin{abstract}
When a vector meson such as $\phi$, $J/\Psi(3S)$ or $\Upsilon(4S)$
decays to a particle-antiparticle pair of neutral mesons, the
time distribution of inclusive decay to a given final state is
naively expected to be the incoherent 
sum of those of the two mesons with
opposite flavors. In this paper, we show that this is
in general not the case for arbitrary coherent two-body states 
of the mesons, and obtain conditions under which such
a naive incoherent sum gives the correct distributions. 
The analysis is based on the Weisskopf-Wigner formalism, and
applicable to the cases where there are more than two orthogonal
states that can mix to form a set of eigenstates of mass and decay
rate.
\end{abstract}



\section{Introduction}
In the studies of $CP$ violation in neutral meson systems
such as $\phi\to K^0\bar K^0$ \cite{Klit},
$J/\Psi(3S) \to D^0\bar D^0$, or 
$\Upsilon(4S)\to B^0\bar B^0$~\cite{NirQuinn},
one often deals with an inclusive decay distribution
where one of the mesons decay to a given final state $f$
at a given time $t$
and the other meson can decay to any final state at any time.
It has been shown in a recent study that
in the decay of $\Upsilon(4S)$, the inclusive decay time distribution of
$\Upsilon(4S)\to f$ is the incoherent 
sum of the decay time distribution of a pure
$B^0$ at $t=0$ decaying to $f$ at time $t$ 
(denoted $\Gamma_{B^0\to f}(t)$) and that of 
$\ovl B^0$ (denoted $\Gamma_{\bar B^0\to f}(t)$)~\cite{lepasym}.
Such relation is critical in analyses of inclusive lepton
asymmetries~\cite{lepasym}, or in relating observed inclusive yield
asymmetries of $\Upsilon(4S)\to f$ and $\Upsilon(4S)\to \bar f$
to the asymmetry of decay amplitudes $Amp(\bar B^0\to f)$
and $Amp(B^0\to \bar f)$~\cite{Kpiasym}.
When a $\Upsilon(4S)$ decays to a pair of neutral $B$ mesons, 
it is generated in a coherent $L=1$ state, which is antisymmetric
under the exchange of the two mesons:
\begin{equation}
   {1\over\sqrt2}
      \left( |\ovl B^0 (\vec k)\ket |B^0 (-\vec k)\ket -
                  |B^0 (\vec k)\ket |\ovl B^0 (-\vec k)\ket \right),
    \label{eq:asymBB}
\end{equation}
where the mesons are labeled by their momentum $\pm \vec k$
which we will drop hereafter and implicitly 
assume that left (right) side of
the meson pair is in $+\vec k (-\vec k)$ direction.
A natural question is then whether such an incoherent sum gives the 
correct inclusive distribution for a general two-body state given by
\begin{equation}
    \Psi =
    a B^0 B^0 + b B^0 \ovl B^0 +
    c \ovl B^0 B^0 + d \ovl B^0 \ovl B^0 \;,
   \label{eq:PsiBBbar}
\end{equation}
where $a,b,c,d$ are arbitrary complex coefficients with
\[
     |a|^2 + |b|^2 + |c|^2 + |d|^2 = 1\;.
\]
When the pair is generated with a definite orbital angular momentum,
further symmetry relations must be satisfied; in this study,
however, we will keep the general form as above.
As we will show
below, the necessary and sufficient condition for the 
naive incoherent sum 
\begin{equation}
  \begin{array}{rl}
   (2|a|^2 + &|b|^2 + |c|^2) \,\Gamma_{B^0\to f}(t) \\
   &+\,(2|d|^2 +  |b|^2 + |c|^2) \,\Gamma_{\bar B^0\to f}(t)
  \end{array}
   \label{eq:inco4S}
\end{equation}
to give the correct distribution for any final state $f$ (and 
independent of the details of the mixing) is
\begin{equation}
     G \eqdef a^* (b+c) + d (b+c)^* = 0\;.
   \label{eq:condBBbar}
\end{equation}

\section{General coherent two-body states}
In the following, we study a system of $n$ orthogonal states
$B_i \; (i=1,\ldots,n)$
mixing to form $n$ eigenstates of mass and decay rate
(physical states)
$B_\alpha \;(\alpha = a,b,c\ldots)$.
The eigenstates $B_\alpha$ are not necessarily orthogonal when
$CP$ is violated.
We use the Weisskopf-Wigner formalism~\cite{Wigner+}, 
but no assumptions
are made on $CP$ or $CPT$ symmetries unless otherwise stated.
The essential approximation used in the formalism is that
the oscillations caused by mass differences and the decay rates
are sufficiently slower than the time scale of decay transitions,
which is a very good assumption for the cases under 
study~\cite{WWdeviation}.

The eigenstates $B_\alpha$ then evolve as
\begin{equation}
    B_\alpha \to e_\alpha(t) B_\alpha \;,\qquad
    e_\alpha(t) \eqdef e^{-(\gamma_\alpha/2 + i m_\alpha) t} \;,
   \label{eq:evolve}
\end{equation}
where $\gamma_\alpha$ and $m_\alpha$ are the decay rate and
mass of the physical state $B_\alpha$. 
The time $t$ is the proper time of the particle under consideration.
The Weisskopf-Wigner formalism can be relativistically
extended to moving particles; it can be
shown, however, that it is equivalent to the evolution in the rest
frame formulated as above~\cite{JSBell}.
We will hereafter consistently
use the indices $i,j$ for the orthogonal states
$B_1$, $B_2 \ldots$, and Greek indexes $\alpha,\beta$ for
the physical states $B_a$, $B_b \ldots\;$:
\[
   \begin{array}{rl}
     \hbox{Orthogoanl states:}\; & 
         i,j,i',j' = 1,2\ldots n,\\ 
     \hbox{Physical states:}\; &
     \alpha,\beta,\alpha',\beta'= a,b
         \ldots\; (n\hbox{ total})\;.
   \end{array}
\]
The eigenstates $B_i$ can be written as linear combination of
$B_\alpha$'s:
\begin{equation}
     B_i = \sum_\alpha r_{i\alpha} B_\alpha\;.
    \label{eq:Bi}
\end{equation}

For the system composed of $B^0$ and $\bar B^0$, we have $n=2$:
\[
    B_1 = B^0 \;, \qquad B_2 = \bar B^0\;,
\]
and the physical states are usually written as
\begin{equation}
   \left\{
   \begin{array}{rcl}
      B_a &=& p  B^0 + q  \ovl B^0 \\
      B_b &=& p' B^0 - q' \ovl B^0 \\
   \end{array} \right.\;,
   \label{eq:Babdef}
\end{equation}
or solving for $B^0$ and $\bar B^0$,
\[
   \left\{
   \begin{array}{rcl}
          B^0 &=& c\, ( q' B_a + q B_b ) \\ 
     \ovl B^0 &=& c\, ( p' B_a - p B_b ) \\ 
   \end{array}
   \right. ,
\]
with
\[
   c \eqdef {1\over p'q + pq'} \;; 
\]
namely,
\begin{equation}
   \begin{array}{ll}
       r_{1a} = cq',& r_{1b} = cq \,,\\
       r_{2a} = cp',& r_{2b} = -cp \,,
   \end{array}
  \label{eq:rpq}
\end{equation}

Returning to the general case of $n$ orthogonal states, 
the orthonormality of $B_i$'s
can be expressed in terms of the physical eigenstates $B_\alpha$ as
\begin{equation}
      \delta_{i\,j}  = \bra B_i|B_j\ket 
         = {\displaystyle \sum_{\alpha\beta} r^*_{i\alpha} r_{j\beta} 
              \,\bra B_\alpha | B_\beta\ket}\;,
   \label{eq:ortho}
\end{equation}
where we have used (\ref{eq:Bi}).
The decay amplitude
of a pure $B_i$ state at $t=0$ decaying to a final state 
$f$ at time $t$ is,
from (\ref{eq:Bi}) and (\ref{eq:evolve}),
\[
   A_{B_i\to f}(t) = 
    \sum_\alpha r_{i\alpha} a_{\alpha f} e_\alpha(t) \; ,
\]
where $a_{\alpha f}$ is the amplitude of $B_\alpha$ decaying to $f$:
\[
    a_{\alpha f} \eqdef Amp(B_\alpha \to f) \;. 
\]
The normalization is such that $|a_{\alpha f}|^2$ is the partial decay
rate of $B_\alpha$ to $f$:
\begin{equation}
     \sum_f |a_{\alpha f}|^2 = \gamma_\alpha \;.
  \label{eq:ampnorm}
\end{equation}
Namely, the density of the final states, 
more precisely the square root of it,
is absorbed into the definition of the amplitude.

The time dependent decay amplitudes $A_{B_i\to f}(t)$ satisfy the
following orthonormality relation \cite{Azimov-ortho}, where
the `inner product' of $A_{B_i\to f}$ and $A_{B_j\to f}$
is defined by integration of $A^*_{B_i\to f}A_{B_j\to f}$ over time
followed by summation over all possible final states:
\begin{equation}
  \begin{array}{rl}
   {\displaystyle
    \sum_f \int_0^\infty} dt & A^*_{B_i\to f}(t) A_{B_j\to f}(t) \\
     = & {\displaystyle 
    \sum_{\alpha\beta}r^*_{i\alpha}r_{j\beta}\sum_f
      a_{\alpha f}^* a_{\beta f}
      \int_0^\infty dt\,e^*_\alpha(t) e_\beta(t) } \\ =
     & {\displaystyle
    \sum_{\alpha\beta}r^*_{i\alpha}r_{j\beta}\;
     {\sum_f a_{\alpha f}^* a_{\beta f}
      \over {\gamma_\alpha + \gamma_\beta \over 2} 
     - i(m_\alpha - m_\beta)} } \\ =
     & \delta_{i\,j}\;.
  \end{array}
  \label{eq:amportho}
\end{equation}
In deriving the above, where we have used the generalized 
Bell-Steinberger relation \cite{Bell-Stein} given by
\begin{equation}
    {\sum_f a_{\alpha f}^* a_{\beta f}
      \over{\gamma_\alpha + \gamma_\beta \over 2} 
     - i(m_\alpha - m_\beta)}
     = \bra B_\alpha | B_\beta\ket\;,
   \label{eq:Bell-Stein}
\end{equation}
together with
the orthonormality of $B_i$'s (\ref{eq:ortho}). Note that the relation
(\ref{eq:Bell-Stein}) reduces to the amplitude normalization condition
(\ref{eq:ampnorm}) for $\alpha=\beta$. 
While the Bell-Steinberger relation
can be derived by requiring that unitarity is satisfied~\cite{Bell-Stein},
it can also be derived from the old-fashioned perturbation
theory to the lowest non-trivial order~\cite{TDLee}.
The probability that a pure $B_i$ at 
$t=0$ decays to a final state $f$ at
time $t$ is simply the square of the time-dependent amplitude:
\[
   \Gamma_{B_i\to f}(t) = |A_{B_i\to f}(t)|^2 \;.
\]
The relation (\ref{eq:amportho}) with $i=j$
shows that this decay distribution conserves probability:
\[
   \sum_f \int_0^\infty dt\,\Gamma_{B_i\to f}(t) = 1 \; .
\]

Now, take a general coherent two-body state at $t=0$ given by
\[
   \Psi(t=0) = \sum_{i\,j} c_{i\,j} B_i B_j \;,
\]
where $c_{i\,j}$ are arbitrary complex coefficients with
\[
    \sum_{i\,j} |c_{i\,j}|^2 = 1\;.
\]
For simplicity, we will hereafter label the left and right sides of 
the particle pair as north (N) and south (S). The specific names to
distinguish the two sides are irrelevant; we just need some labels for the
two orthogonal spaces.
The probability that north side decays to a final state $f_N$ at time $t_N$
and the south side to a final state $f_S$ at time $t_S$ is then
\begin{equation}
    \Gamma_{\Psi\to f_N f_S}(t_N,t_S) =
     \Big| \sum_{i\,j} c_{i\,j} A_{B_i\to f_N}(t_N) A_{B_j\to f_S}(t_S)
     \Big|^2\;.
  \label{eq:double-t}
\end{equation}
From the orthonormality of the decay amplitude (\ref{eq:amportho}), one sees
that this double-time decay distribution also conserves probability; namely,
when integrated over the two decay times and summed over all possible 
final states, it becomes unity:
\begin{equation}
    \sum_{f_N\,f_S} \int_0^\infty dt_N \int_0^\infty dt_S\;
       \Gamma_{\Psi\to f_N f_S}(t_N,t_S) = 1\;.
    \label{eq:2tnorm}
\end{equation}
We now define the inclusive decay distribution of $\Psi$ to a final state
$f$, where $f$ can come from either side of the decay:
\begin{equation}
   \begin{array}{rl}
    \Gamma_{\Psi\to f}(t) \eqdef &
      {\dspl \sum_{f_N} \int_0^\infty dt_N 
      \Gamma_{\Psi\to f_N f}(t_N,t) } \\ 
      +& {\dspl \sum_{f_S} \int_0^\infty dt_S \;
       \Gamma_{\Psi\to f f_S}(t,t_S)}\;,
   \end{array}
   \label{eq:incdef}
\end{equation}
which, due to (\ref{eq:2tnorm}), satisfies
\[
   \sum_f \int_0^\infty dt\,\Gamma_{\Psi\to f}(t) = 2\;.
\]
The number 2 comes from the fact that the final state $f$ can come
from either side of the decay.
The question is under what condition this is 
equal to the naive incoherent sum
\[
  \begin{array}{rl}
    \Gamma^{\rm naive}_{\Psi\to f}(t) &\eqdef {\displaystyle 
   \sum_{i\,j} |c_{i\,j}|^2 \left( \Gamma_{B_i\to f}(t) + 
                                   \Gamma_{B_j\to f}(t) \right) }\\ 
    &= {\displaystyle 
   \sum_{i\,j}\, \Big(\, |c_{i\,j}|^2 
     + |c_{j\,i}|^2\Big)\, \Gamma_{B_i\to f}(t) } \;,
  \end{array}
\]
which is the generalization of (\ref{eq:inco4S}). 
Using the expression of the
double-time distribution (\ref{eq:double-t}) 
and the orthonormality of the
decay amplitude (\ref{eq:amportho}), the inclusive decay distribution
(\ref{eq:incdef}) becomes
\[
   \begin{array}{rl}
    \Gamma_{\Psi\to f}(t) & = {\dspl
       \sum_{i\, i' j} (c^*_{i\,j} c_{i'j} + c^*_{ji} c_{ji'})
               A^*_{B_i\to f}(t) A_{B_{i'}\to f}(t) }             \\
                           = & {\dspl
       \Gamma^{\rm naive}_{\Psi\to f}(t) } \\ + & {\dspl
       \sum_{i \not= i'} \sum_j(c^*_{i\,j} c_{i'j} + c^*_{ji} c_{ji'})
               A^*_{B_i\to f}(t) A_{B_{i'}\to f}(t) }\;.
   \end{array}       
\]
The necessary and sufficient condition for this to be equal to
$\Gamma^{\rm naive}_{\Psi\to f}(t)$ is then
\begin{equation}
  \sum_{i \not= i'}\, G_{ii'}\,
               A^*_{B_i\to f}(t) A_{B_{i'}\to f}(t)  = 0\;.
    \label{eq:condition}
\end{equation}
with
\begin{equation}
     G_{ii'} \eqdef \sum_j(c^*_{i\,j} c_{i'j} + c^*_{ji} c_{ji'}) \;.
\end{equation}
A sufficient condition for (\ref{eq:condition})
to be satisfied is clearly
\begin{equation}
    G_{ii'} = 0
    \quad (\hbox{for all } i\not= i') .
   \label{eq:gencond}
\end{equation}
The matrix $G_{ii'}$ is `hermitian' in the sense that
\[
    G_{ii'} = G_{i'i}^*\;,
\]
which guarantees that $\Gamma_{\Psi\to f}(t)$ is a real quantity.
Note also that the norm of $G_{ii'}$ is re-phase invariant; 
namely, when the
phase of $B_i$'s are re-defined, $G_{ii'}$ simply changes its phase:
\[
 B_i\to B_i\, e^{i\phi_i}\qquad\longrightarrow\qquad
 G_{ii'}\to  G_{ii'}\,e^{i(\phi_i-\phi_{i'})}\;.
\]
Thus, the condition (\ref{eq:gencond}) is re-phase invariant.

In the case of $n=2$, the condition~(\ref{eq:gencond}) 
becomes Eq.~(\ref{eq:condBBbar}) with
$G = G_{12}$.
In this case, it is straightforward to show that the
condition $G=0$ is the necessary as well as sufficient
condition as long as $\gamma_a \not= \gamma_b$, $m_a \not= m_b$,
and coefficients $p,q,p',q'$ are all non-zero. We still require
that $\Gamma^{\rm naive}_{\Psi\to f}(t)$ is correct independent of
decay amplitudes $a_{\alpha f}$.
The proof for general case is given in the appendix. The
derivation is simple if $CPT$ is conserved in the
mixing; namely, $p'=p$ and $q'=q$. Then the 
condition (\ref{eq:condition}) becomes
\begin{equation}
  \begin{array}{l}
     \Re(Gpq^*) \left( e^{-\gamma_a t}|a_{af}|^2 -
                        e^{-\gamma_b t}|a_{bf}|^2\right) \\
         + 2\, \Im(Gpq^*) \;
      \Im\left( e^{-(\gamma_+ - i\delta m)t} a_{af}^* a_{bf}\right) = 0\;,
  \end{array}
  \label{eq:condCPT}      
\end{equation}
with
\[
   \gamma_+ \equiv {\gamma_a + \gamma_b\over 2},\quad 
   \delta m \equiv m_a - m_b\;.
\]
The three terms in
({\ref{eq:condCPT}) have different time dependences, and thus 
each term should be separately zero.
When $\gamma_a \not= \gamma_b$, this condition is equivalent to
$\Re(Gpq^*)=\Im(Gpq^*)=0$, or simply
\[
    Gpq^* = 0 \qquad (CPT)\;.
\]
Thus, if both $p$ and $q$ are non-zero (i.e. there is 
a mixing), $G$ must be zero in order for the naive incoherent sum
to be correct.

Let's briefly appreciate the meaning of the condition $G=0$.
This is satisfied by any coherent state of
$B^0\bar B^0$ and $\bar B^0 B^0$ since then we have
$a = d = 0$. It includes the $\Upsilon(4S)$ case given by
(\ref{eq:asymBB}), or a
$B^0\bar B^0$ state with an even orbital angular momentum.
Also, the condition is satisfied if $b+c=0$ regardless of
the values of $a$ and $d$. However, the condition is not
satisfied, for example,
by the symmetric state
\[
    B^0 B^0 + B^0 \ovl B^0 +
    \ovl B^0 B^0 + \ovl B^0 \ovl B^0\;.
\]
Such a state cannot be readily produced in practice, 
but in principle it is
possible if there exists an interaction with $\Delta B$ = 2, 
such as the hypothetical superweak interaction. 

To summarize, we have studied inclusive decay time distributions of
coherent two-body states. 
We find that there is a set of orthonormality 
relations among decay time distributions of states that are pure
and orthogonal to each other at $t=0$. Using this, we have shown
that the naive incoherent sum of single particle decay time
distributions does not always 
give the correct inclusive distribution,
and extracted conditions for it to be the case. Such incoherent sum
was found to be correct for any $B^0\bar B^0$ state regardless of
the orbital angular momentum. 

\acknowledgements}
I would like to thank Y. Azimov, R. Briere, I. Dunietz, 
S. Glashow, S. Pakvasa,
for useful discussions, and R. Madrak for reading the manuscript.
This work was supported by the Department of Energy
Grant DE-FG02-91ER40654.

\vskip0.2in
\appendix
\section*{}
For $n=2$, the condition (\ref{eq:condition}) reduces to
\begin{equation}
  \begin{array}{l}
   \Re(Gq'^*p')\, e^{-\gamma_a t} |a_{af}|^2 - 
   \Re(Gq^*p)\, e^{-\gamma_b t} |a_{bf}|^2  \\ + 
   \Re\left[ (G^*p'^*q - Gq'^*p)\, 
       e^{-(\gamma_+ - i\delta m)t} a_{af}^* a_{bf}\right] = 0\;.
  \end{array}
  \label{eq:condgenBB}
\end{equation}
Since the three terms have different time dependences, each term should
separately be zero. The decay amplitudes are in general
non-zero (we are requiring that the naive inclusive
distribution be correct independent of decay amplitudes); thus,
the above condition leads to
\begin{equation}
  \left\{
  \begin{array}{rl}
     \Re(Gq'^*p') &= 0 \\
     \Re(Gq^*p) &=0 \\
     G^*p'^*q - Gq'^*p &= 0 
  \end{array} \right. \;,
  \label{eq:cond1}
\end{equation}
where the last condition is due to the fact that the term
$e^{i \delta m t}$ samples all possible phases.
Now we define
\begin{equation}
  G \eqdef |G|e^{ig},\quad s \eqdef q e^{ig},\quad s' \eqdef q' e^{ig}.
  \label{eq:sdef}
\end{equation}
Substituting this in (\ref{eq:cond1}), and dividing each
equation by $|s|^2$, one obtains
\begin{equation}
  \left\{
  \begin{array}{rl}
    {\dspl |G|\,\Re\left({p'\over s'}\right)} &= 0 \\
    {\dspl |G|\,\Re\left({p\over s}\right)} &=0 \\
    {\dspl |G|\,\left[ \left({p'\over s'}\right)^* - {p\over s} 
                \right] } &= 0 
  \end{array} \right. \; ;
  \label{eq:cond1}
\end{equation}
namely,
\[
   |G|=0,\quad {\rm or}\quad \left\{
     \begin{array}{l}
       {\dspl {p'\over s'},\; {p\over s}}: \hbox{pure imaginary, and} \\
       {\dspl \left({p'\over s'}\right)^* = {p\over s}}
     \end{array} \right.
\]
Thus, if $|G|\not= 0$,this leads to
\begin{equation}
   {p'\over -q'} = {p\over q}\;,
\end{equation}
which means that $B_a$ and $B_b$ are same
physical states and it contradicts our assumption that they have different
decay rates and masses. Thus, $G=0$ results from (\ref{eq:condgenBB}).
On the other hand, the condition $G=0$ trivially leads to
(\ref{eq:condgenBB}); 
thus, $G=0$ is the necessary and sufficient condition for
the naive inclusive distribution to be correct assuming that
$\gamma_a\not=\gamma_b$, $m_a\not= m_b$ and $p,q,p',q'$ are
non-zero (namely, $B^0$ and $\bar B^0$ mix).


\vspace{1cm}

\begin{references}
\bibitem{Klit}
  See for example,
  C. D. Buchanan, R. Cousins, C. Dib, R.D. Peccei, and J. Quackenbush,
  Phys. Rev. {\bf D45}, 4088 (1992); 
  M. Hayakawa and A.I. Sanda, Phys. Rev. {\bf D48}, 1150 (1993).
\bibitem{NirQuinn}
  Y. Nir and H.R. Quinn, in `$B$ Decays', 2nd ed., 
  Ed. S. Stone, World Scientific, 1994, and references therein.
\bibitem{lepasym}
  H. Yamamoto, hep-ph/9703336, to be published in Phys. Lett. {\bf B}.
\bibitem{Kpiasym}
  For example, if one assumes that 
  $Amp(B^0\to K^-\pi^+)$ = $Amp(\bar B^0\to K^+\pi^-)$ = 0, then
  in spite of $B^0$-$\bar B^0$ mixing or $CP$ violation, one
  can show that the inclusive yield ratio measured on $\Upsilon(4S)$
  is related to the amplitude ratio by $N(K^-\pi^+)/N(K^+\pi^-)$
  = $|Amp(\bar B^0\to K^-\pi^+)/Amp(B^0\to K^+\pi^-)|^2$.
\bibitem{Wigner+}
  V.F. Weisskopf and E.P. Wigner, Z. Phys. {\bf 63}, 54 (1930); 
  {\it ibid.}, {\bf 65}, 18 (1930).
\bibitem{WWdeviation}
  Typical deviations in decay distributions from the Weisskopf-Wigner
  approximation is of order $\delta m/m\ll 1$. For more discussions,
  see for example, P.K. Kabir and A. Pilafsis, Phys. Rev. {\bf A53}, 66 (1996),
  and references therein.
\bibitem{JSBell}
  J.S. Bell, {\it Theory of Weak Interactions} in 
  Les Houches Summer School on Theoretical Physics, Les
  Houches, France, 1965, ed. C. de Witt and M. Jacob.
\bibitem{Bell-Stein}
  The relation can readily be obtained from $\sum_f |Amp(\phi\to f)|^2 = 
  - d/dt \bra \phi | \phi \ket$, where $\phi$ is an arbitrary linear
  combination of $B_i$.
  J.S. Bell and J. Steinberger, in Proceedings of the Oxford International
  Conference on Elementary Particles, 1965.
\bibitem{TDLee}
  T.D. Lee, {\it Particle Physics and Introduction to Field Theory},
  Harwood Academic Publishers, 1981
\bibitem{Azimov-ortho}
  I would like to thank Y. Azimov for pointing out this interpretation.
\end{references}
\end{document}